\documentclass[manuscript, letterpaper]{aastex}
\usepackage{amsbsy}
\usepackage{amsfonts}
\usepackage{amssymb}

\usepackage{graphicx}
\usepackage{graphics}
\usepackage{bm}
\begin{document}


\title{Observational Quantification of the Energy Dissipated \\ by Alfv\'en Waves in a Polar Coronal Hole: \\ Evidence that Waves Drive the Fast Solar Wind.}
\author{M. Hahn\altaffilmark{1} and D. W. Savin\altaffilmark{1}}

\altaffiltext{1}{Columbia Astrophysics Laboratory, Columbia University, MC 5247, 550 West 120th Street, New York, NY 10027 USA}

\date{\today}

\begin{abstract}
	
	We present a measurement of the energy carried and dissipated by Alfv\'en waves in a polar coronal hole. Alfv\'en waves have been proposed as the energy source that heats the corona and drives the solar wind. Previous work has shown that line widths decrease with height in coronal holes, which is a signature of wave damping, but have been unable to quantify the energy lost by the waves. This is because line widths depend on both the non-thermal velocity $v_{\mathrm{nt}}$ and the ion temperature $T_{\mathrm{i}}$. We have implemented a means to separate the $T_{\mathrm{i}}$ and $v_{\mathrm{nt}}$ contributions using the observation that at low heights the waves are undamped and the ion temperatures do not change with height. This enables us to determine the amount of energy carried by the waves at low heights, which is proportional to $v_{\mathrm{nt}}$. We find the initial energy flux density present was $6.7 \pm 0.7 \times 10^{5}$~$\mathrm{erg\,cm^{-2}\,s^{-1}}$, which is sufficient to heat the coronal hole and accelerate the solar wind during the 2007 - 2009 solar minimum. Additionally, we find that about 85\% of this energy is dissipated below 1.5~$R_{\sun}$, sufficiently low that thermal conduction can transport the energy throughout the coronal hole, heating it and driving the fast solar wind. The remaining energy is roughly consistent with what models show is needed to provide the extended heating above the sonic point for the fast solar wind. We have also studied $T_{\mathrm{i}}$, which we found to be in the range of 1 - 2~MK, depending on the ion species. 
	
\end{abstract}

\maketitle
	
\section{Introduction} 

	One of the major models to describe the heating of the solar corona and the acceleration of the solar wind relies on waves to carry the energy. Such wave-driven models have been supported by observations that waves, and in particular Alfv\'enic waves \citep{VanDoorsselaere:ApJ:2008,Goossens:AA:2009}, are observed throughout the solar atmosphere from the chromosphere \citep{Jess:Sci:2009,Depontieu:Sci:2007,McIntosh:Nature:2011}, to the corona \citep{Tomczyk:Sci:2007}, and into the solar wind \citep{Belcher:JGR:1971}. 
	
	However, one difficulty for simple wave-driven models has been that Alfv\'en waves are predicted to dissipate via viscosity, thermal conductivity, and resistivity relatively far from the Sun, at about 2 - 5~$R_{\sun}$ \citep[e.g.,][]{Parker:ApJ:1991, Cranmer:SSR:2002, Ofman:SSR:2005, Ofman:LRSP:2010}. In order for waves to heat the corona they must be damped at much lower heights where heat conduction is more efficient. For this reason theories have been developed for how the waves may dissipate more quickly. These theories rely on the inhomogeneity of the corona and show, for example, that the waves can be more strongly damped through phase mixing \citep{Heyvaerts:AA:1983, Ofman:ApJ:2002}, turbulent cascade \citep{Matthaeus:ApJ:1999}, resonant absorption \citep{Goossens:SSR:2011}, or the nonlinear generation of compressive waves and shocks \citep{Ofman:ApJ:1997, Ofman:ApJ:1997a, Suzuki:ApJ:2005}. 

	Only recently has there been found clear observational evidence for dissipation of Alfv\'en waves at low heights on open field lines. \citet{Hahn:ApJ:2012} and \citet{Bemporad:ApJ:2012} studied coronal hole observations and demonstrated that Alfv\'en waves are damped at relatively low heights in the corona. In these studies the Alfv\'en waves were observed spectroscopically through the non-thermal broadening of optically thin spectral lines. The magnitude of the non-thermal broadening is predicted to be proportional to the wave amplitude \citep{Banerjee:AA:1998,Doyle:SolPhys:1998,Moran:ApJ:2003,Banerjee:AA:2009}; and for energy to be conserved the wave amplitude must increase with height above the Sun as the density decreases \citep{Hollweg:SolPhys:1978, Moran:AA:2001}.  However, \citet{Hahn:ApJ:2012} and \citet{Bemporad:ApJ:2012} found that the line widths decrease above about $1.2$~$R_{\sun}$. They ruled out systematic errors as the cause of the decrease. This confirmed earlier indications that the line widths decrease at these heights \citep{Banerjee:AA:1998, Doyle:AA:1999, Moran:ApJ:2003, Oshea:AA:2005, Dolla:AA:2008}. 

	In order to determine if waves are indeed responsible for heating the corona and driving the solar wind one must quantify both the energy initially present in the waves as well as that dissipated by the waves. This has been difficult to determine because the measurement of a line width includes contributions both from thermal broadening, which is proportional to the ion temperature $T_{\mathrm{i}}$, and from non-thermal broadening, which is proportional to unresolved plasma motions along the line of sight. A conventional method to estimate the non-thermal velocity $v_{\mathrm{nt}}$ is to assume some value for $T_{\mathrm{i}}$. For example, \citet{Bemporad:ApJ:2012} assumed the ion temperature was equal to the ionization equilibrium formation temperature of the ion emitting the line. However, this is not necessarily correct since some studies have shown that $T_{\mathrm{i}}$ may be much greater than the formation temperature  \citep[e.g.,][]{Tu:ApJ:1998,Landi:ApJ:2009,Hahn:ApJ:2010}. Because $T_{\mathrm{i}}$ is expected to be larger than $T_{\mathrm{e}}$,  \citet{Hahn:ApJ:2012} assumed that $T_{\mathrm{i}} \geq T_{\mathrm{e}}$ and thereby estimated an upper bound for $v_{\mathrm{nt}}$. 
		
	Here we present a method to separately determine $v_{\mathrm{nt}}$ and $T_{\mathrm{i}}$. The data are described in Section~\ref{sec:obs} and the analysis method is presented in Section~\ref{sec:anal}. The $v_{\mathrm{nt}}$ results are given in Section~\ref{sec:vnt}. The non-thermal velocity is proportional to the wave amplitudes and from $v_{\mathrm{nt}}$ we can determine the initial wave energy, the change in the wave energy flux density, and the length and time scales over which the waves are damped. These data indicate that waves are sufficient to heat the coronal hole and drive the fast solar wind. They also provide quantitative constraints for theoretical models of wave damping. In Section~\ref{sec:ti} we present our measurements of ion temperatures and compare them to some earlier measurements that found only lower and upper bounds for the temperature. These temperature data can be used to test various models for ion heating in the corona. In Section~\ref{sec:scat} we consider possible systematic errors from instrument scattered light and show that they do not significantly affect the analysis. We summarize our results in Section~\ref{sec:sum}. 

\section{Observation}\label{sec:obs}

	Our data come from four observations made with the Extreme ultraviolet Imaging Spectrometer \citep[EIS;][]{Culhane:SolPhys:2007} on \emph{Hinode} \citep{Kosugi:SolPhys:2007}. The observations were made on 2009 April 23 at 12:42, 13:16, 13:50, and 15:17. Each observation was 30 minutes in duration. For these data the 2$^{\prime\prime}$ slit was pointed at a polar coronal hole at positions relative to the central meridian of $X = -14.5^{\prime\prime}$, $15.5^{\prime\prime}$, $45.4^{\prime\prime}$, and 105.6$^{\prime\prime}$, respectively. The height range covered by the slit extended from about 0.95~$R_{\sun}$ to about $1.45$~$R_{\sun}$. These data are the same as used by \citet{Hahn:ApJ:2012}, but excluding their observation centered at $X=-44.5^{\prime\prime}$. We omitted that particular observation as it had a density scale height at low heights that was larger compared to the other pointings, possibly due to intervening quiet Sun material. Our results here for wave damping, though, are consistent with the previous results that included the additional observation. 
	
	The four pointings were averaged together in order to improve the statistical accuracy. This was done by first using the standard EIS processing routines to clean the data of spikes, warm pixels, and dark current, and calibrate the data. Drifts in the wavelength scale were then corrected using the method described by \citet{Kamio:SolPhys:2010}. After aligning the data to the same wavelength scale, pixels at the same radius from each of the four observations were averaged to create the dataset analyzed. Finally, these data were further binned in the vertical direction. For the analysis described below, where we perform a fit to the data at low heights, we have used a binning of 8 pixels per bin ($\sim 0.01$~$R_{\sun}$). To extend these results to larger heights, where the intensities are correspondingly much smaller, we have used a 32 pixel binning ($\sim 0.03$~$R_{\sun}$).

\section{Analysis Method}\label{sec:anal}
	
\subsection{Line Widths}\label{subsec:width}	

	We fit Gaussian functions to the spectrum in order to derive the line widths. In particular, each line was fit with a double Gaussian so as to account for both actual off-disk emission and the instrument scattered light. The lines used for various aspects of the analysis are given in Table~\ref{table:linelist}. The fitting procedures are described in detail in \citet{Hahn:ApJ:2012}. Here we only briefly review the method.
	
	Instrument scattered light is expected to superimpose the spectrum of the solar disk emission onto the off-disk data. Because line widths tend to be narrower on the disk, scattered light can be a significant source of systematic error at large heights when the fraction of real emission is small. To correct for the scattered light we first measured line profiles from the portions of our observation that looked at the solar disk. Then we constructed a predicted scattered light line profile for each line. For these parameters we used the measured line width and centroid position and 2\% of the on-disk intensity. This last value is based on estimates for the magnitude of the scattered light in EIS \citep{Hahn:ApJ:2012}. Then, for each position in the off-disk data, a fit was performed using a Gaussian with free parameters added to the artificial scattered light profile. This is equivalent to subtracting the scattered light profile from the spectrum. We include in the analysis only data where the the fraction of the total intensity due to scattered light is less than 45\%, since below this limit the results are insensitive to the precise amount of stray light. The characterization of the scattered light and its possible systematic effects on the analysis are described in detail in \citet{Hahn:ApJ:2012} and in Section~\ref{sec:scat}.
	
	To evalute the uncertainties of the fitted parameters, we used the same Monte Carlo type of uncertainty analysis as described in \citet{Hahn:ApJ:2012}. That is, we first fit the original data. Then we added random numbers to each data point, where the distribution of these random numbers was chosen to have a standard deviation equal to the residual between each point and the initial fit. These modified data were fit and the process was repeated several hundred times. The uncertainties on the fit parameters are given by the standard deviation of the results from the many fits. We used this approach, rather than taking the least squares fit uncertainties derived from the initial fit, because it takes into account possible systematic errors when the fitting function is not a perfect representation of the data. For example, weak features or unflagged warm pixels are treated as noise by this analysis, which is reflected in the uncertainties. 
	
		The measured full width at half maximum $\Delta \lambda_{\mathrm{FWHM}}$ of an optically thin spectral line depends on instrumental broadening $\Delta \lambda_{\mathrm{inst}}$, the ion temperature $T_{\mathrm{i}}$, and the non-thermal velocity $v_{\mathrm{nt}}$ as \citep{Ultraviolet}  
\begin{equation}
\Delta \lambda_{\mathrm{FWHM}} = \left[ \Delta\lambda_{\mathrm{inst}}^2 +  
4 \ln(2)\left(\frac{\lambda}{c}\right)^{2}\left(\frac{2k_{\mathrm{B}}T_{\mathrm{i}}}{M} + v_{\mathrm{nt}}^2 \right) \right]^{1/2}. 
\label{eq:width}
\end{equation}
Here $\lambda$ is the wavelength of the line, $c$ is the speed of light, $k_{\mathrm{B}}$ is the Boltzmann constant, and $M$ is the mass of the ion. We have subtracted the instrumental width using the $\Delta \lambda_{\mathrm{inst}}$ values as a function of position along the slit tabulated by \citet{Young:EIS:2011}. These data for $\Delta \lambda_{\mathrm{inst}}$ are also supported by an independent calibration by \citet{Hara:ApJ:2011} who compared an EIS observation to visible line spectra. The instrumental FWHM is about 0.06~\AA\ and the typical thermal plus non-thermal FWHM is about 0.04 -- 0.06~\AA. After subtracting the instrumental width, the observed width can then be expressed as an effective velocity, 
\begin{equation}
v_{\mathrm{eff}}=\sqrt{v_{\mathrm{th}}^2 + v_{\mathrm{nt}}^2}, 
\label{eq:veffdefine}
\end{equation}
where $v_{\mathrm{th}} = \sqrt{2k_{\mathrm{B}}T_{\mathrm{i}}/M}$. This $v_{\mathrm{eff}}$ depends on both $T_{\mathrm{i}}$ and $v_{\mathrm{nt}}$. 
	
\subsection{Separating Thermal and Non-thermal Broadening}\label{subsec:method}
	
	\citet{Dolla:AA:2008} pointed out that the thermal and non-thermal contributions can be inferred if two assumptions are made. One can then calculate $v_{\mathrm{nt}}(R_{0})$ at a radius $R_{0}$ using data from another height $R_{1}$. The first assumption is that $v_{\mathrm{th}}$ is constant with height for each ion emitting the line being studied. This implies that
\begin{equation}
v_{\mathrm{eff}}^2(R_{1})- v_{\mathrm{eff}}^2(R_{0}) = v_{\mathrm{nt}}^2(R_1) - v_{\mathrm{nt}}^2(R_0).
\label{eq:assumpt1}
\end{equation}
The other assumption is that the waves are undamped. By conservation of energy, this implies that $v_{\mathrm{nt}} \propto n_{\mathrm{e}}^{-1/4}$ \citep[see e.g.,][]{Hollweg:SolPhys:1978, Moran:AA:2001}. Since the waves are assumed to be undamped we have 
\begin{equation}
\frac{v_{\mathrm{nt}}(R_1)}{v_{\mathrm{nt}}(R_0)} = \left[\frac{n_{\mathrm{e}}(R_1)}{n_{\mathrm{e}}(R_0)}\right]^{-1/4}.
\label{eq:assumpt2}
\end{equation}
Putting it all together, one finds 
\begin{equation}
v_{\mathrm{nt}}(R_0) = \left\{\frac{v_{\mathrm{eff}}^2(R_1)-v_{\mathrm{eff}}^2(R_0)}{\left[\frac{n_{\mathrm{e}}(R_1)}{n_{\mathrm{e}}(R_0)}\right]^{-1/2} -1}\right\}^{1/2}.
\label{eq:DSvnt}
\end{equation}
\citet{Dolla:AA:2008} used this method to determine $v_{\mathrm{nt}}(R_0)$ by taking an average over the results for a fixed $R_0$ while varying $R_1$. However, they did not observe damping and they inferred a quite small $v_{\mathrm{nt}}$, possibly for the reasons we discuss below in Section~\ref{sec:vnt}. 

	The method we use relies on the same assumptions as the \citet{Dolla:AA:2008} method, but the application is somewhat different. Here, we use a least squares fit. The reason for doing this is that uncertainties in the data can cause large variations in the $v_{\mathrm{nt}}(R_0)$ determined using equation~(\ref{eq:DSvnt}). A least squares fit implicitly takes these uncertainties into account and is more robust to noise. Combining equations~(\ref{eq:veffdefine}), (\ref{eq:assumpt1}), and (\ref{eq:assumpt2}), the function used in the fit is
\begin{equation}
v_{\mathrm{eff}}(R) = \sqrt{v_{\mathrm{th}}^2 + v_{\mathrm{nt}}^2(R_0) \left[\frac{n_{\mathrm{e}}(R)}{n_{\mathrm{e}}(R_0)}\right]^{-1/2}}. 
\label{eq:ls1}
\end{equation}
Here, $v_{\mathrm{th}}$ and $v_{\mathrm{nt}}(R_0)$ are the only free parameters to be determined. As we discuss later, we assume that $v_{\mathrm{th}}$ for a given ion is the same for every height, though it can be different for each ion. Solving equations~(\ref{eq:DSvnt}) or (\ref{eq:ls1}) requires the ratio $n_{\mathrm{e}}(R)/n_{\mathrm{e}}(R_0)$. We describe below how this is determined.

	The assumption of constant ion temperature is reasonable for low heights. Ions in coronal holes are known to be heated, although the precise mechanism has not been determined. Some possibilities include ion cyclotron resonance heating by high frequency waves \citep{Cranmer:SSR:2002} and stochastic heating by turbulent fluctuations that disturb the ion orbits \citep{Chandran:ApJ:2010}. In both cases the heating rate is predicted to depend on the charge to mass ratio, with minor ions heated more strongly than protons. UVCS measurements have shown that the proton temperature is roughly constant from about 1.3~$R_{\Sun}$ to 2~$R_{\sun}$ with a temperature of 1 -- 2~$\times 10^{6}$~K \citep{Esser:ApJ:1999}. Indirect measurements have inferred a proton temperature of about $1.8\times10^{6}$~K at the base of the corona \citep{Hahn:ApJ:2013}. Both these measurements suggest that the proton temperature is constant at low heights in the corona. For this reason, Coulomb collisions with the protons are expected to cool the minor ions and maintain them at a relatively steady temperature at low heights \citep{Landi:ApJ:2009}. Measurements estimating upper and lower bounds for $T_{\mathrm{i}}$ at heights of $R \lesssim 1.15$~$R_{\sun}$ have shown that $T_{\mathrm{i}}$ is consistent with being constant over this height range, albeit with large uncertainties \citep{Landi:ApJ:2009,Hahn:ApJ:2010}. Note again that each ion may have a different $T_{\mathrm{i}}$ (i.e., $v_{\mathrm{th}}$) which we assume does not change with height.
	
	It is also reasonable to assume that waves are undamped at low heights, and consequently $v_{\mathrm{nt}} \propto n_{\mathrm{e}}^{-1/4}$. This theoretical relation is valid for outward propagating waves when the solar wind velocity is much smaller than the Alfv\'en speed \citep{Cranmer:ApJS:2005}, a condition expected to be met at low heights. Numerous studies have observed the predicted trend for $R \lesssim 1.15$~$R_{\sun}$ \citep{Doyle:SolPhys:1998, Banerjee:AA:1998, Banerjee:AA:2009,Hahn:ApJ:2012}. In estimating $v_{\mathrm{nt}}$, these studies have assumed $T_{\mathrm{i}}$ to be either the ion formation temperature or the electron temperature, but have found the same $n_{\mathrm{e}}^{-1/4}$ trend. Thus, this trend is not very sensitive to uncertainties in the magnitude of $T_{\mathrm{i}}$. 
	
	Based on the above, in the range 1.02 - 1.12~$R_{\sun}$ the ion temperatures should be reasonably constant with height; although, $T_{\mathrm{i}}$ may still differ depending on the ion species. Also, the upper height of $1.12$~$R_{\sun}$ is below the point where the waves appear to be damped. Thus, it is reasonable to perform the fits to equation~(\ref{eq:ls1}) over these heights.
	
\begin{figure}
\centering \includegraphics[width=0.9\textwidth]{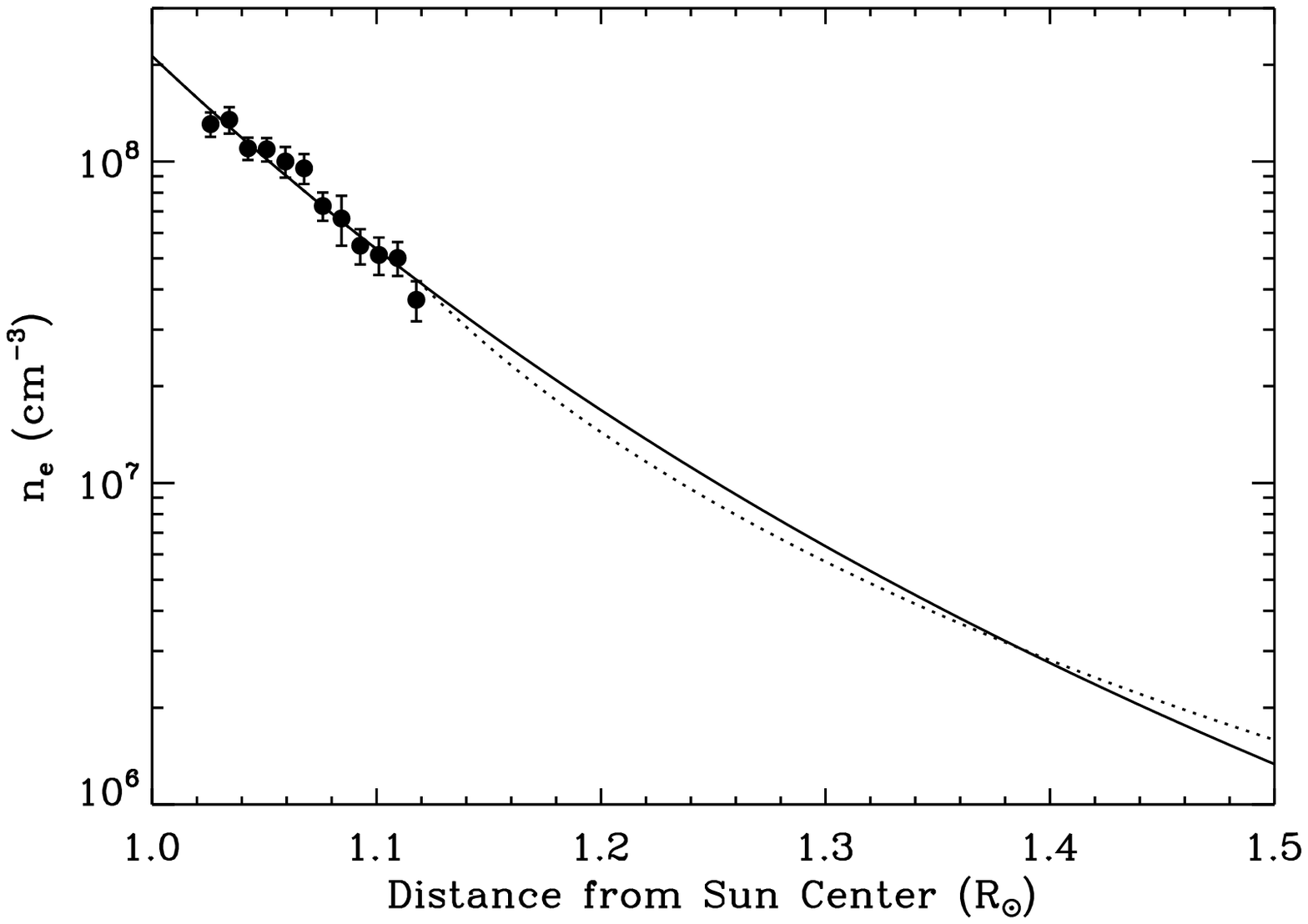}
\caption{\label{fig:density} The filled circles indicate the electron density $n_{\mathrm{e}}$ derived from an Fe~\textsc{ix} intensity ratio. The solid line shows the hydrostatic equilibrium fit to the data using equation~(\ref{eq:hydrodensity}) in the range $1.02$ - $1.12$~$R_{\sun}$. The dotted line shows the empirical model from \citet{Cranmer:ApJS:2005}, scaled to match the data at 1.12~$R_{\sun}$. For the analysis we used the hydrostatic fit for $R < 1.12$~$R_{\sun}$ and extended to larger heights using the empirical model.
}
\end{figure}

	Our analysis also requires an independent measurement of $n_{\mathrm{e}}$. This was obtained from the intensity ratio of the Fe~\textsc{ix} 188.50~\AA\ and 189.94~\AA\ lines using atomic data from CHIANTI \citep{Dere:AA:1997,Landi:ApJ:2012}. Figure~\ref{fig:density} shows the inferred densities, which are typical of densities found in other coronal hole observations \citep{Wilhelm:AAR:2011}. At these low heights the solar wind velocity is small and the corona is close to hydrostatic equilibrium. We therefore fit the density over the range 1.02 - 1.12~$R_{\sun}$ using \citep[e.g.,][]{Guhathakurta:ApJ:1992, Doyle:AA:1999}
\begin{equation}
n_{\mathrm{e}}(R)=n_{\mathrm{e}}(R_0)\exp\left[\frac{-(R-R_0)}{HR_{0}R}\right],
\label{eq:hydrodensity}
\end{equation}
where all the lengths are measured in units of $R_{\sun}$ and $H$ is the density scale height, which was found to be $H=0.0657\pm0.0052$~$R_{\sun}$. Here and throughout we give all uncertainties at a 1$\sigma$ statistical confidence level. The fit is illustrated in figure~\ref{fig:density}. Using this expression for the density, the ratio $n_{\mathrm{e}}(R)/n_{\mathrm{e}}(R_0)$ in equation~\ref{eq:ls1} can be rewritten so that 
\begin{equation}
v_{\mathrm{eff}}(R) = \sqrt{v_{\mathrm{th}}^2 + v_{\mathrm{nt}}^2(R_0) \left[\exp\frac{-(R-R_0)}{RR_0H}\right]^{-1/2}}.
\label{eq:ls2}
\end{equation}
This function is useful for our analysis as some of the statistical fluctuations in the magnitude of $n_{\mathrm{e}}$ are smoothed out, while retaining the essential description of the density variation with height. 

	Since the height variation for $v_{\mathrm{th}}$ and $v_{\mathrm{nt}}$ are completely determined by our assumptions and the measured $n_{\mathrm{e}}$, the results are independent of the particular value of $R_{0}$ chosen for the analysis. Here, we used $R_{0} = 1.05$~$R_{\sun}$. 

\subsection{Uncertainties}\label{subsec:uncert}
	
\begin{figure}
\centering \includegraphics[width=0.9\textwidth]{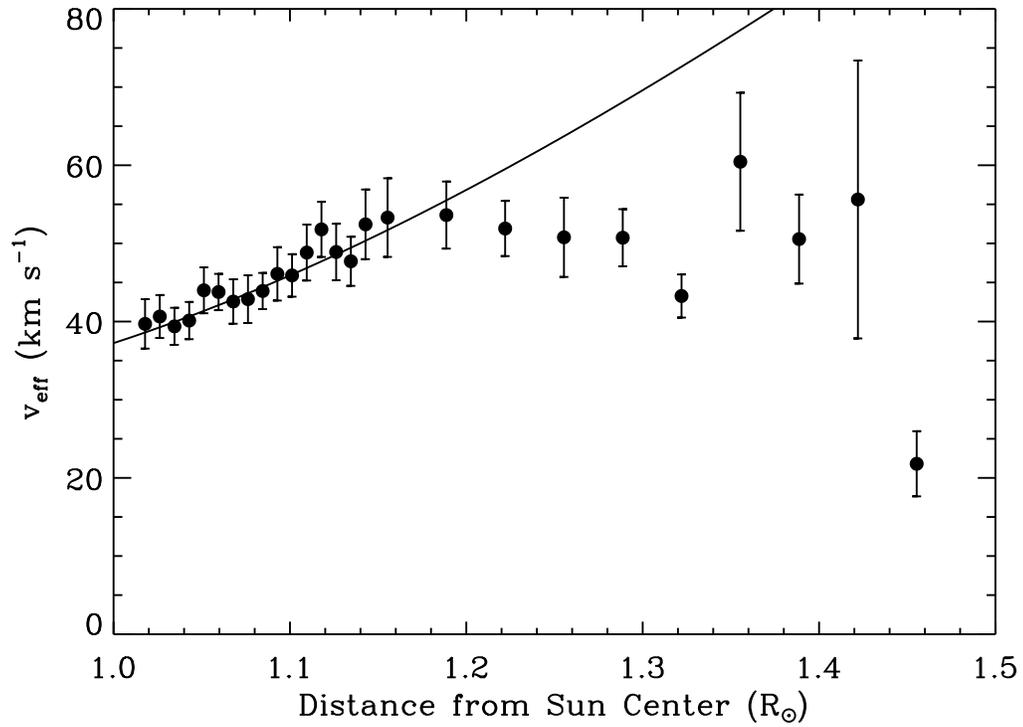}
\caption{\label{fig:veff_fit} The filled circles show the measured effective velocity $v_{\mathrm{eff}}$ for Fe~\textsc{xi}~188.22~\AA. The solid line illustrates the average fit to the data between 1.02 and 1.12~$R_{\sun}$ using equation~(\ref{eq:ls2}). The fit parameters for this ion were $v_{\mathrm{th}} = 25.8 \pm 5.4$~$\mathrm{km\,s^{-1}}$ and $v_{\mathrm{nt}}=32.2 \pm 4.2$~$\mathrm{km\,s^{-1}}$. 
}
\end{figure}

	Uncertainties in the line widths $v_{\mathrm{eff}}$ and $H$ were propagated into the fitted parameters $v_{\mathrm{th}}$ and $v_{\mathrm{nt}}$ using a Monte Carlo analysis. The observed data for each fit are the single inferred value of $H$ and a set of values for $v_{\mathrm{eff}}$ as a function of height for a given ion. Normally distributed random variations were added to these input data. The magnitude of these variations was set so that the standard deviation of the random numbers added was equal to the $1\sigma$ uncertainty for each data point used in the fit of equation~(\ref{eq:ls2}) over the height range 1.02 - 1.12~$R_{\sun}$ (e.g., Figure~\ref{fig:veff_fit}). The fits were then performed for 1000 different variations, with each iteration producing different values of $v_{\mathrm{th}}$ and $v_{\mathrm{nt}}$. Next we took the mean of each parameter and estimated the $1\sigma$ scatter to be the standard deviation. 
	
	We found that for some of the iterations, the value of $v_{\mathrm{th}}$ for a given ion would imply $T_{\mathrm{i}} < T_{\mathrm{e}}$. Although this is clearly a possible fit to the data, the result does not seem physically reasonable. At very low heights, below the range of our fit, the density is high enough that electrons, protons, and ions should be in equilibrium so that $T_{\mathrm{i}}=T_{\mathrm{e}}$. At large heights the ions are observed to be heated and $T_{\mathrm{i}} > T_{\mathrm{e}}$ and collisions can be neglected. In the range where we perform the fits, the situation lies in-between these two extremes, that is the ions are expected to be heated in some way, but also to be cooled by collisions with protons having $T_{\mathrm{p}} \geq T_{\mathrm{e}}$ \citep{Esser:ApJ:1999,Landi:ApJ:2009,Hahn:ApJ:2013}. Thus, throughout the observed height range we expect $T_{\mathrm{i}} \geq T_{\mathrm{e}}$. We have applied this constraint to our analysis by rejecting fits that imply $T_{\mathrm{i}} < 8\times10^{5}$~K, which is consistent with a previous analysis of this observation \citep{Hahn:ApJ:2012} and is a typical $T_{\mathrm{e}}$ for a coronal hole \citep{Wilhelm:AAR:2011}.  	

\section{Wave Amplitude and Energy}\label{sec:vnt}	

	Fits to equation~(\ref{eq:ls2}) were performed over the range 1.02 - 1.12~$R_{\sun}$. The fits used a total of eleven lines from five ions (see column three of Table~\ref{table:linelist}). Lines formed at higher temperatures, such as Fe~\textsc{xii} and Fe~\textsc{xiii}, were not considered, because a differential emission measure analysis showed that they come from warmer structures \citep{Hahn:ApJ:2012}. Figure~\ref{fig:veff_fit} shows an example of the fit for Fe~\textsc{xi}. If all the lines originate from the same volume, then it is expected that they will all have the same $v_{\mathrm{nt}}$, though not necessarily the same $v_{\mathrm{th}}$. As expected, the inferred $v_{\mathrm{nt}}$ from the five different ions were in reasonable agreement with one another (Table~\ref{table:fitresults}). Thus, we took the unweighted mean of the results from the different ions to find that $v_{\mathrm{nt}} = 33.0 \pm 2.4$~$\mathrm{km\,s^{-1}}$ at 1.05~$R_{\sun}$. 
	
	In the study of \citet{Dolla:AA:2008} they found $v_{\mathrm{nt}} = 15 \pm 2$~$\mathrm{km\,s^{-1}}$, which is significantly smaller than we find here. There are several possible explanations for this apparent discrepancy. \citet{Dolla:AA:2008} focussed on an observation made in May 2002. This time period was near solar maximum, whereas our data were obtained near solar minimum and so the difference may reflect some solar cycle variation. Additionally, they described the polar coronal hole as ``not well developed'' and so their data likely contains other structures along the line of sight. For their analysis they studied a line from Mg~\textsc{x}, which is a lithium-like ion with a peak formation $T_{\mathrm{e}} \approx 1.3 \times 10^{6}$~K, but also has a tail of high ion abundance towards much greater temperatures \citep{Bryans:ApJ:2009}. Thus, if $v_{\mathrm{nt}}$ is smaller in hotter structures than it is in a coronal hole, then their value would be systematically underestimated. 
	
\begin{figure}
\centering \includegraphics[width=0.9\textwidth]{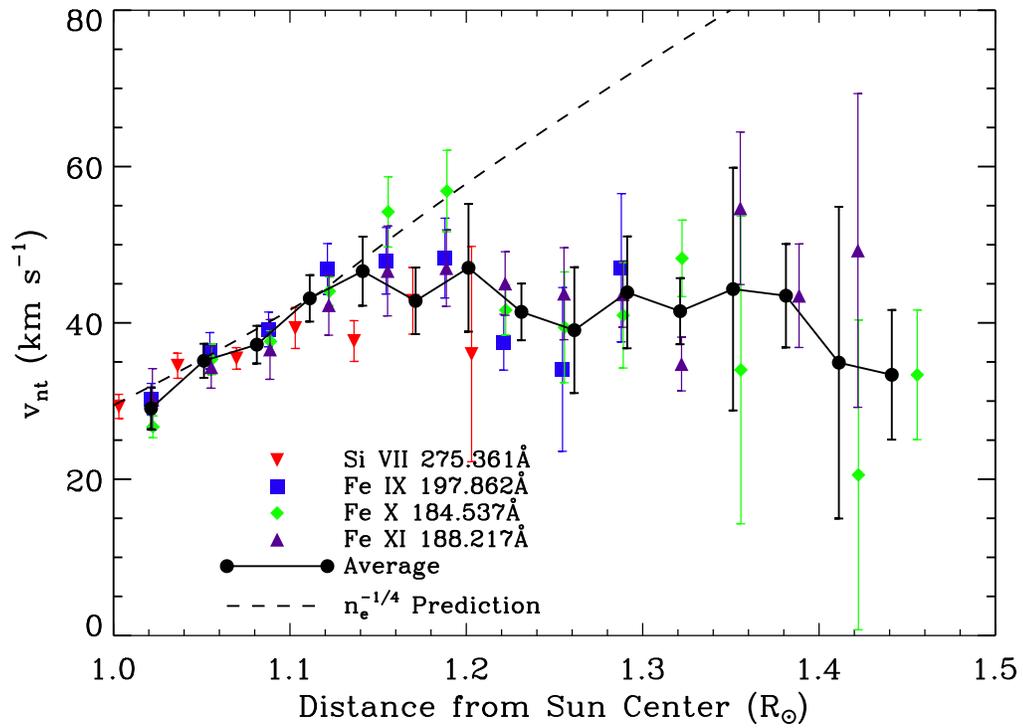}
\caption{\label{fig:vnt} Symbols indicate the non-thermal velocity $v_{\mathrm{nt}}$ from the strongest observed lines. The filled circles and solid line show the averaged $v_{\mathrm{nt}}$ combining the data from the various ions. The dashed line illustrates the predicted electron density $n_{\mathrm{e}}^{-1/4}$ trend for undamped waves. 
}
\end{figure}

	Figure~\ref{fig:vnt} shows $v_{\mathrm{nt}}$ as a function of $R$ for Si~\textsc{vii}~275.37~\AA, Fe~\textsc{ix}~197.86~\AA, Fe~\textsc{x}~184.54~\AA, and Fe~\textsc{xi}~188.22~\AA. These lines, listed in the fourth column of Table~\ref{table:linelist}, were chosen because they could be observed to relatively large heights. In each case the corresponding $v_{\mathrm{th}}$ from Table~\ref{table:fitresults} has been subtracted from $v_{\mathrm{eff}}$ using the values for each ion determined from the fits. The solid line in the figure shows the unweighted mean $v_{\mathrm{nt}}(R)$ in 0.03~$R_{\sun}$ bins for these lines and the dashed line shows the predicted $n_{\mathrm{e}}^{-1/4}$ trend for undamped waves. The data show that $v_{\mathrm{nt}}$ is consistent with undamped waves below about $1.15$~$R_{\sun}$. We also find that the $v_{\mathrm{nt}}$ derived from each ion species is the same, which justifies the assumption that all ions experience the same fluid motions. At larger heights, we find that $v_{\mathrm{nt}}$ deviates from the $n_{\mathrm{e}}^{-1/4}$ trend, which implies wave damping. 
	
	The energy density flux carried by the waves can be estimated using \citep{Doyle:SolPhys:1998, Banerjee:AA:1998, Moran:AA:2001} $$F = 2\rho v_{\mathrm{nt}}^2 V_{\mathrm{A}},$$ where $\rho$ is the mass density and 
$$V_{\mathrm{A}}=B/\sqrt{4\pi\rho}$$ 
is the Alfv\'en speed with $B$ being the magnetic field strength. To estimate the varying magnetic field strength for the superradially expanding polar coronal hole we used the empirical model from equation~(6) of \citet{Cranmer:ApJ:1999}. In terms of the area expansion $A(R)/A(R_{\sun})$ this gives $$B(R)=B(R_{\sun})A(R_{\sun})/A(R).$$ The polar magnetic field can vary by a few Gauss between solar cycles and has smaller variations within a solar minimum \citep{Wang:ApJ:2009}. \citet{Wang:ApJ:2010} gives a median $B(R_{\sun})=7.3$~G for this solar minimum with a spread of $\sim 1$~G. At low heights $\rho$ can be found from the measured $n_{\mathrm{e}}$. For larger heights it was necessary to extrapolate the density measurements. We did this using the profile from \citet{Cranmer:ApJS:2005} which was based on white light measurements out to several $R_{\sun}$. Their $n_{\mathrm{e}}(R)$ function was scaled to match our measurements at 1.12~$R_{\sun}$. The uncertainty of the scaling factor was taken to match that of $n_{\mathrm{e}}$ at 1.12~$R_{\sun}$. Additionally, we found that $F$ remains nearly identical if we simply use the hydrostatic fit for $n_{\mathrm{e}}$ over the entire height range. 
	
\begin{figure}
\centering \includegraphics[width=0.9\textwidth]{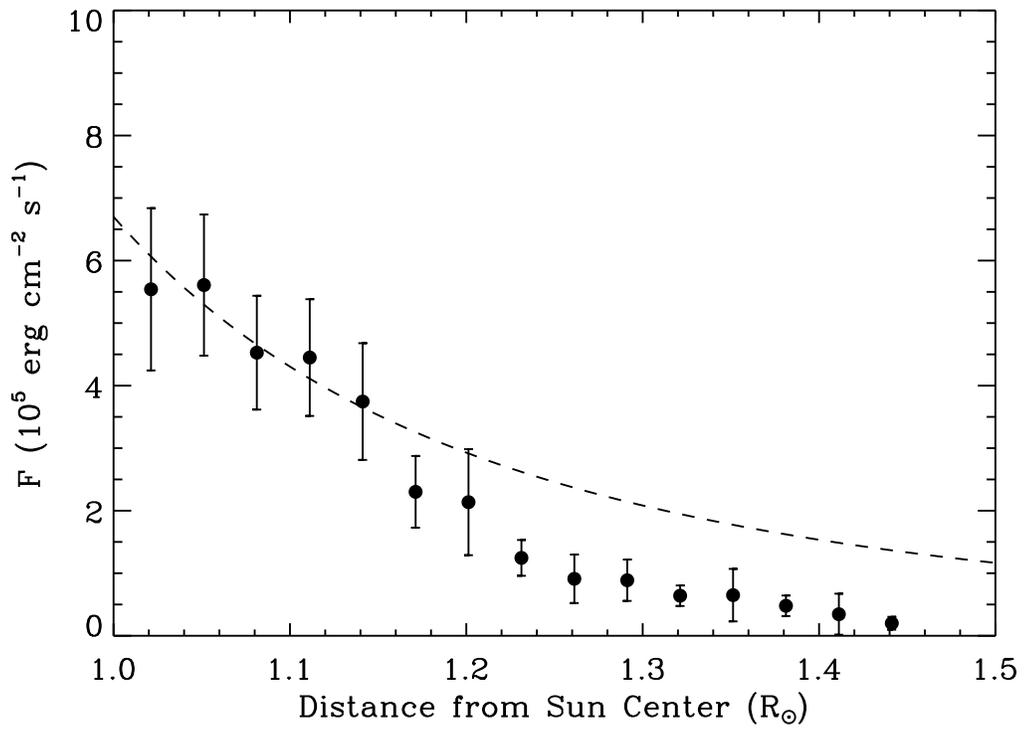}
\caption{\label{fig:F} Wave energy density flux $F$ as a function of height (filled circles). The dashed line illustrates the predicted trend for undamped waves. 
}
\end{figure}

	Figure~\ref{fig:F} shows the energy density flux $F$ as a function of height based on the averaged results for $v_{\mathrm{nt}}$, plotted in Figure~\ref{fig:vnt}. These data are also listed in Table~\ref{table:ergresults}. The error bars represent the combined uncertainties from $v_{\mathrm{nt}}$, $\rho$ and $B$. One can see that $F$ is decreasing with height, but some of this decrease is due simply to the expansion of the coronal hole. The dashed line in Figure~\ref{fig:F} illustrates the variation of $F$ with height for undamped waves, where the decrease is due only to the superradial expansion of the coronal hole \citep{Cranmer:ApJ:1999}. Clearly the waves are damped more rapidly with height than predicted by expansion alone. 
	
\begin{figure}
\centering \includegraphics[width=0.9\textwidth]{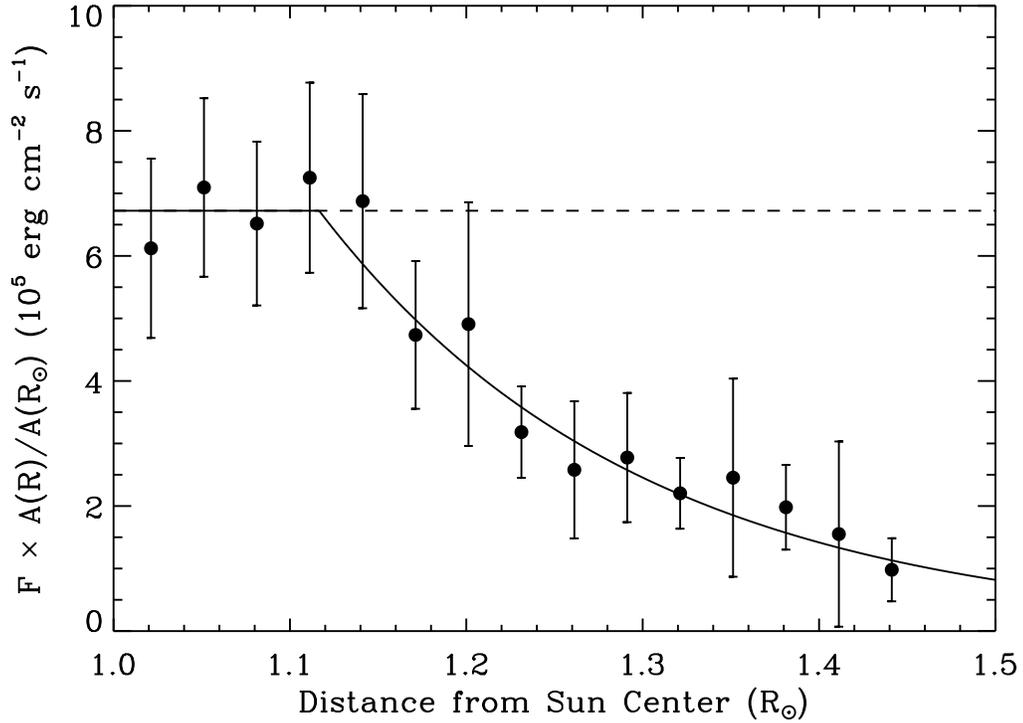}
\caption{\label{fig:FA} Points indicate the wave energy density flux $F$ multiplied by the expansion factor $A(R)/A(R_{\sun})$ in order to more clearly show the decrease due to damping. For undamped waves, the points would fall on a horizontal line. The dashed line indicates the average of the points below 1.12~$R_{\sun}$. This average shows that $F = 6.7 \pm 0.7 \times10^{5}$~$\mathrm{erg\,cm^{-2}\,s^{-1}}$ is present in the waves at 1~$R_{\sun}$. The solid line gives an exponential fit, from which a damping length of $0.18\pm0.04$~$R_{\sun}$ was derived. The point at which the exponential decay begins was a free parameter of the fit, with the result $R=1.12\pm0.04$~$R_{\sun}$.
}
\end{figure}

	To more clearly show the effect of damping we show the quantity $F A(R)/A(R_{\sun})$ in Figure~\ref{fig:FA} (also listed in Table~\ref{table:ergresults}). In this plot, measurements for undamped waves would fall on a horizontal line. The data are consistent with undamped waves for $R \lesssim 1.15$~$R_{\sun}$. The dashed line in Figure~\ref{fig:FA} is drawn at the average of the points below $1.12$~$R_{\sun}$, which is $F = 6.7 \pm 0.7 \times 10^{5}$~$\mathrm{erg\,cm^{-2}\,s^{-1}}$. This is the amount of Alfv\'en wave energy present at the base of the corona. \citet{Withbroe:ARAA:1977} estimated that $8\times10^{5}$~$\mathrm{erg\,cm^{-2}\,s^{-1}}$ is typically required to heat a coronal hole and accelerate the fast solar wind. About $7\times10^{5}$~$\mathrm{erg\,cm^{-2}\,s^{-1}}$ goes into driving the solar wind, and the rest of the energy is lost through radiation and conduction. However, during the 2007 - 2009 solar minimum the solar wind was observed to be unusually weak, being slower, less dense, and cooler than during the previous minimum \citep{McComas:GRL:2008, Wang:ApJ:2010}. The solar wind power was about 25\% less, while other conditions in coronal holes remained similar \citep{Hahn:ApJ:2010}. 
This implies that for the recent solar minima only roughly $5\times10^{5}$~$\mathrm{erg\,cm^{-2}\,s^{-1}}$ would be required to drive the solar wind. So, after including radiation and conduction, the total coronal hole energy requirement is $\sim6\times10^{5}$~$\mathrm{erg\,cm^{-2}\,s^{-1}}$. Our measurements show not just that the amount of energy carried by the waves is sufficient to account for coronal heating and solar wind acceleration within the coronal hole. They also indicate that the waves are indeed damped, with $F A(R)/A(R_{\sun})$ falling from about $6.7\times10^{5}$~$\mathrm{erg\,cm^{-2}\,s^{-1}}$ at $1$~$R_{\sun}$ to $1\times10^{5}$~$\mathrm{erg\,cm^{-2}\,s^{-1}}$ by 1.44~$R_{\sun}$. Thus, the waves lose $\sim85\%$ of their initial energy by 1.44~$R_{\sun}$. These findings indicate that the waves do in fact provide most of the required heating. 
	
	The length and time scales over which the waves are damped provide benchmarks for theoretical calculations \citep[e.g.,][]{Zaqarashvili:AA:2006,Pascoe:AA:2012}. In order to estimate the length scale over which the waves are damped we fit an exponential to $F A(R)/A(R_{\sun})$. This fit is illustrated by the solid line in Figure~\ref{fig:FA}. The initial value of $F$, the height where damping begins $R_{\mathrm{d}}$, and the exponential damping length $L_{\mathrm{d}}$, were free parameters of the fit. The initial $F$ was the same as found above, $F = 6.7 \pm 0.7 \times10^{5}$~$\mathrm{erg\,cm^{-2}\,s^{-1}}$. The fit yielded $R_{\mathrm{d}} = 1.12 \pm 0.04$~$R_{\sun}$, which is consistent with the point where $v_{\mathrm{nt}}$ deviates from the $n_{\mathrm{e}}^{-1/4}$ trend (e.g., Figure~\ref{fig:vnt}). The relatively large error bar is due to the coarse binning used here. We find that the damping length is about $L_{\mathrm{d}} = 0.18 \pm 0.04$~$R_{\sun}$. This is significantly shorter than the ad hoc heating scale length currently used in coronal heating models. For example, \citet{Downs:ApJ:2010} assumed a heating scale height of $0.7$~$R_{\sun}$ for coronal holes. We can also estimate a timescale for the damping. This was done by converting distance $R$ to wave travel time $t$ using the fact that the velocity of the waves is about the Alfv\'en speed, which varies from about 1 -- 2~$\times 10^{3}$~$\mathrm{km\,s^{-1}}$ over the height range of this observation. Taking $V_{\mathrm{A}}(R)$ into account and fitting an exponential to the data as a function of $t$, we find that the damping time is about $68 \pm 15$~s. This damping time is of of similar magnitude or slightly shorter than the expected wave periods. 

	Solar wind models show that in order to accelerate the fast solar wind to the speeds observed far from the Sun, some input of wave energy is needed above the point where the solar wind becomes supersonic \citep{Cranmer:SSR:2002}. This suggests that not all of the wave energy should be damped at very low heights. The amount of initial energy that is required to be undamped to large heights is about $1\times10^{5}$~$\mathrm{erg\,cm^{-2}\,s^{-1}}$. Our results show that at least up to about $1.4$~$R_{\sun}$ sufficient energy remains in the waves to provide the additional acceleration for the solar wind. However, because of the large uncertainties, our data can also be consistent with the wave energy going to zero at large distances. To get a rough estimate we performed a similar fit to the one described in the above paragraph, but using an exponential plus a constant. We find that at large distances $F A(R)/A(R_{\sun}) \rightarrow 0.6 \pm 1.4 \times 10^{5}$~$\mathrm{erg\,cm^{-2}\,s^{-1}}$. One additional source of uncertainty for this estimate is that we do not know how the ion temperatures are changing with height. As the dissipation of the wave energy is likely to heat the ions, the assumption of constant temperature probably becomes less reasonable at the larger heights in our observation. Since any resulting thermal broadening would increase with height, our assumption of constant $v_{\mathrm{th}}$ would cause us to overestimate $v_{\mathrm{nt}}$, underestimate the change in $v_{\mathrm{nt}}$ with height, and thereby underestimate the actual damping. However, increasing $T_{\mathrm{i}}$ would decrease the wave energy available for the extended solar wind acceleration while the assumption that $T_{\mathrm{i}}$ is constant over the observed heights allows for a reasonable partition of the energy deposition between the low and extended corona.

\section{Ion Temperatures}\label{sec:ti}
	
\begin{figure}
\centering \includegraphics[width=0.9\textwidth]{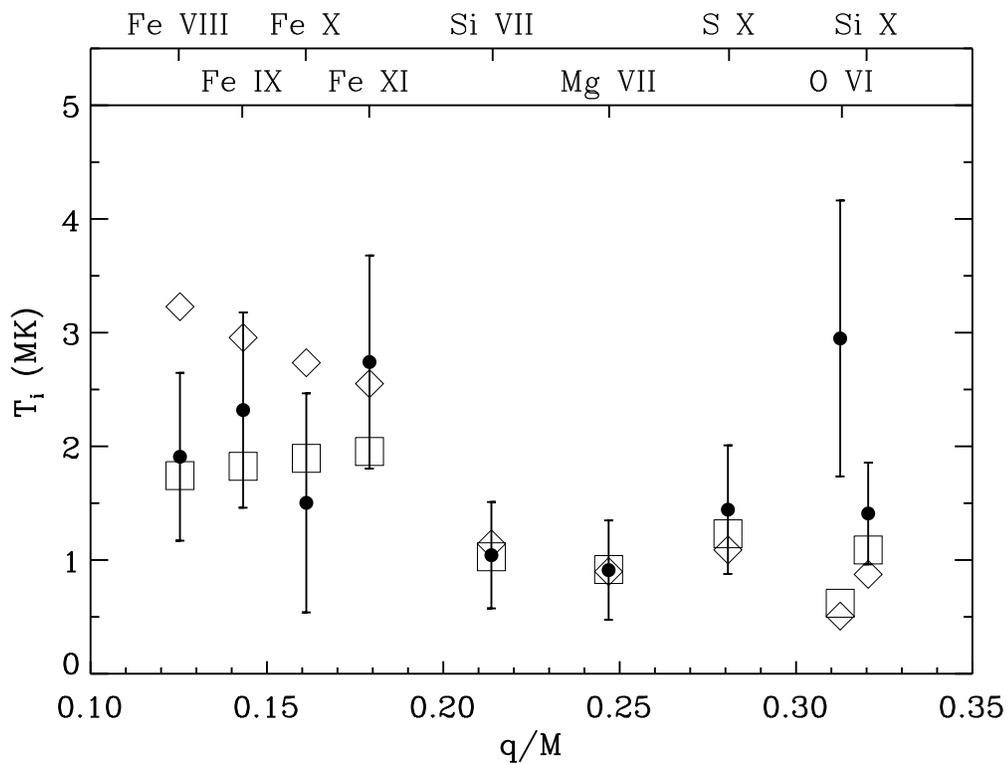}
\caption{\label{fig:Ti} Filled circles show the temperature $T_{\mathrm{i}}$ for each ion, derived by subtracting the average $v_{\mathrm{nt}} = 33.0 \pm 2.4$~$\mathrm{km\,s^{-1}}$ from $v_{\mathrm{eff}}$, plotted versus charge to mass ratio $q/M$ for different ion species. The open squares and open diamonds show the pattern of $T_{\mathrm{i}}$ versus $q/M$ predicted by the models of \citet{Cranmer:ApJ:1999a} and  \citet{Chandran:ApJ:2010}, respectively. 
}
\end{figure}

	The temperature of each ion can be inferred at 1.05~$R_{\sun}$ from all the line widths observed at that height (column five of Table~\ref{table:linelist}) by subtracting the non-thermal width $v_{\mathrm{nt}} = 33.0 \pm 2.4$~$\mathrm{km\,s^{-1}}$. The circles in Figure~\ref{fig:Ti} shows $T_{\mathrm{i}}$ for each of the ions measured as a function of charge to mass ratio $q/M$, in units of elementary charge $e$ per atomic mass unit (amu). These data are also given in Table~\ref{table:Tiresults}. 
	
	 Previous measurements have found that $T_{\mathrm{i}}$ is  greater than $T_{\mathrm{e}}$ for $q/M \lesssim 0.2$, while for slightly higher $q/M$ ions $T_{\mathrm{i}} \approx T_{\mathrm{e}}$, but it may increase again for $q/M \gtrsim 0.3$ \citep{Landi:ApJ:2009,Hahn:ApJ:2010}. Here we find a similar pattern with respect to $q/M$. We find that for $q/M < 0.2$, $T_{\mathrm{i}}\approx 2 \times 10^{6}$~K. For larger $q/M$, $T_{\mathrm{i}} \approx 1 \times 10^{6}$~K, which is about the expected value of $T_{\mathrm{e}}$ for a coronal hole. At even higher $q/M > 0.3$ there is a suggestion that $T_{\mathrm{i}}$ increases based on the S~\textsc{x} and O~\textsc{vi} data. The yet higher $q/M$ point from Si~\textsc{x} appears to contradict this trend. However, there are systematic uncertainties for Si~\textsc{x} and S~\textsc{x} because both ions are formed at relatively high temperatures and so a large fraction of the emission may come from structures outside the coronal hole \citep{Hahn:ApJ:2012}. Ions formed at even higher temperatures, such as Fe~\textsc{xii} and Fe~\textsc{xiii} were omitted from the analysis because most of the emission in those lines comes from plasma with $\log T_{\mathrm{e}} > 6.1$, and so probably does not come from the same structure as the rest of our data. We should also note that our uncertainties are large enough that we cannot rule out that $T_{\mathrm{i}}$ is actually constant over the entire range with respect to $q/M$.

	We have observed the effects of low frequency non-resonant waves on the measured line width. However, theories to explain the observed properties of $T_{\mathrm{i}}$ rely on turbulence and high frequency resonant waves. Such waves can be generated by a turbulent cascade, which transports some of the energy in the low frequency waves to high frequencies \citep{Matthaeus:ApJ:1999}. 
	
	The specific ion heating mechanism may be due to resonant interactions between the ions and ion cyclotron waves \citep{Cranmer:ApJ:1999a, Isenberg:ApJ:2007} or through stochastic heating by the turbulence \citep{Chandran:ApJ:2010}. These different models for ion heating predict different dependences of $T_{\mathrm{i}}$ on $q/M$. Thus, $T_{\mathrm{i}}$ measurements can be used to test these models. For example, \citet{Cranmer:ApJ:1999a} developed a model in which the ions are heated by ion cyclotron waves. In order to make use of equations (2) and (15) of \citet{Cranmer:ApJ:1999a}, we ignore collisions and assume a typical solar wind plasma wave spectral index of $3/2$ \citep{Leamon:JGR:1998, Podesta:ApJ:2007, Chandran:ApJ:2010}. Then one finds
\begin{equation}
T_{\mathrm{i}} \propto M \left(\frac{q}{M}\right)^{1/2}\left(1-\frac{q}{M}\right). 
\label{eq:cranmerti}
\end{equation}
In the model of \citet{Chandran:ApJ:2010}, Alfv\'en wave turbulence causes ion orbits to become stochastic and absorb energy from the turbulence. They derive a dependence of $T_{\mathrm{i}}$ on $q/M$. Using their model, which ignores collisions, and if we also assume that (a) the ratio of the turbulent velocity fluctuations to the thermal velocity perpendicular to the magnetic field are the same for all the ions and (b) the turbulent fluctuations have spectral index $3/2$, then using equation~(21) of \citet[][]{Chandran:ApJ:2010} we find that 
\begin{equation}
T_{\mathrm{i}} \propto M\left(\frac{M}{q}\right)^{2/3}. 
\label{eq:chandranti}
\end{equation}
The neglect of collisions in deriving either of these trends is probably not a very good approximation at these low heights. Nevertheless, we can compare these predictions to our data. The open squares and diamonds in Figure~\ref{fig:Ti} illustrate the predicted trends from \citet{Cranmer:ApJ:1999a} and \citet{Chandran:ApJ:2010}, respectively. In each case the theoretical trends have been multiplied by a scaling factor that was chosen to produce the best average agreement with the observations. Given the large uncertainties in our analysis and the neglect of collisions in the models, both the ion cyclotron resonance heating and stochastic heating by Alfv\'en wave turbulence models show reasonable qualitative agreement with our data.

\section{Scattered Light} \label{sec:scat}

	Instrumental scattered light has been a major source of systematic uncertainty for previous measurements of line widths in the solar corona. Such stray light is expected to superimpose an unshifted solar disk spectrum onto the off-disk data. Since the on-disk line widths are narrower, contamination by scattered light tends also to make the off-disk data narrower. We corrected for this effect by subtracting a scattered light line profile from our data using the methods described by \citet{Hahn:ApJ:2012}. In specific, we measured the line width, centroid position, and intensity for each line at the lowest available on-disk position in our data, which was about 0.95~$R_{\sun}$. We then fit the off-disk data with a double Gaussian profile, one Gaussian having free parameters and the other having fixed parameters derived from the on-disk measurements. For the fixed parameters we used the measured line width and centroid position. We took the stray light intensity to be 2\% of the on-disk intensity. 
 	
 	 In reality, the 2\% estimate for the scattered light relative to the disk intensity is an upper limit, based on measurements of line intensities. \citet{Hahn:ApJ:2012} showed that the intensity of the He~\textsc{ii} line falls below 2\% of the on-disk intensity for heights greater than about 1.15~$R_{\sun}$. Since some of the observed He~\textsc{ii} intensity is due to real emission, the stray light fraction must in fact be less than 2\%. 
 	
 	Additional support for this result can be found from the intensity of other lines. For this we have measured the intensity of the oxygen lines O~\textsc{iv} 279.94~\AA\ and 279.63~\AA, O~\textsc{v} 248.46~\AA, and O~\textsc{vi} 183.94~\AA\ and 184.12~\AA. These lines are formed at relatively cool temperatures of $\log T_{\mathrm{e}}(\mathrm{K}) = $ 5.2, 5.4, and 5.5, for O~\textsc{iv}, \textsc{v}, and \textsc{vi}, respectively \citep{Bryans:ApJ:2009}. For this reason, they are expected to be present in the transition region and visible in the on-disk data, but should be weak in the off-disk data, which does not look into the transition region.
 	
\begin{figure}
\centering \includegraphics[width=0.9\textwidth]{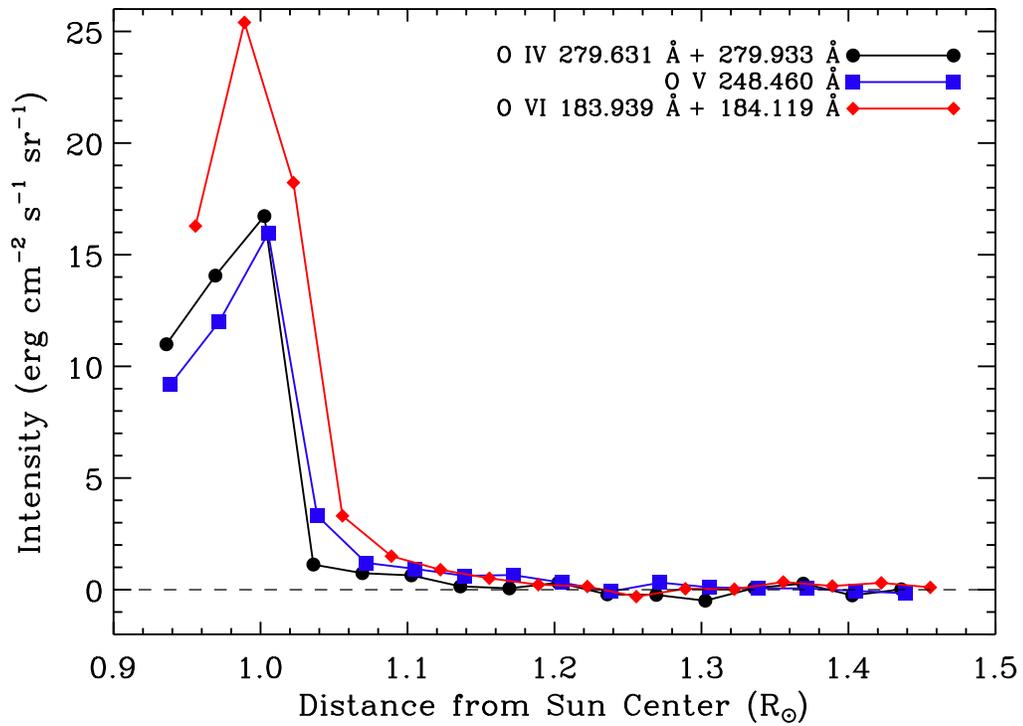}
\caption{\label{fig:oxygen} Intensity versus height for lines from O~\textsc{iv}, \textsc{v}, and \textsc{vi}. The dashed line on the plot is drawn at zero intensity. Based on the average of the intensities above 1.20~$R_{\sun}$, the stray light level relative to the lowest on-disk point is $-0.006 \pm 0.025$ for O~\textsc{iv}, $0.008 \pm 0.019$ for O~\textsc{v}, and $0.006 \pm 0.012$ for O~\textsc{vi}. See text for details. 
}
\end{figure}

 	For each of these oxygen lines, Figure~\ref{fig:oxygen} shows the intensity versus height. Here, no scattered light subtraction has been performed. Because these lines become weak in the off-disk data, it was not possible to determine the intensity using the usual method of fitting the line profiles to a Gaussian. Instead, the intensity was measured by integrating the spectrum over a wavelength range containing the lines. The background intensity was determined from the average of the points at the limits of the integration. The figure shows that the intensity drops off rapidly with height, becoming essentially zero by about 1.2~$R_{\sun}$. We can use these profiles to estimate the scattered light fraction relative to the intensity of the lowest observed on-disk point. Taking the average of the intensities above $1.20$~$R_{\sun}$ we find this fraction is $-0.006 \pm 0.025$ for O~\textsc{iv}, $0.008 \pm 0.019$ for O~\textsc{v}, and $0.006 \pm 0.012$ for O~\textsc{vi}. Thus, the scattered light level is consistent with zero based on these lines. 

\begin{figure}
\centering \includegraphics[width=0.9\textwidth]{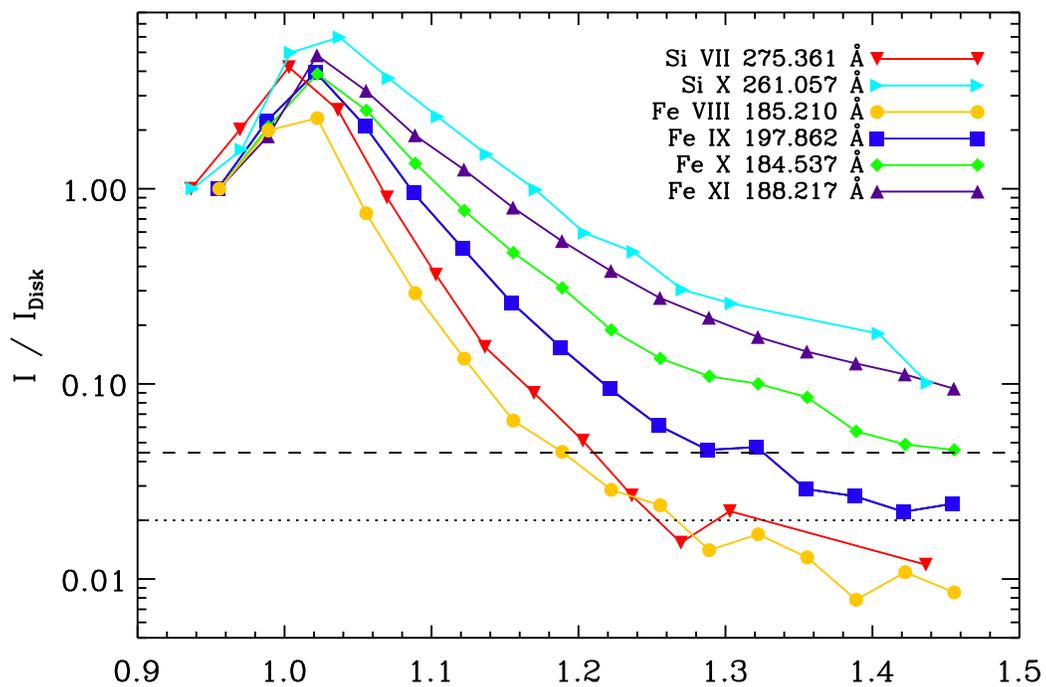}
\caption{\label{fig:ivsr} Intensity normalized to the on-disk intensity at about 0.95~$R_{\sun}$ for lines used in the analysis plus lines from Fe~\textsc{viii} and Si~\textsc{x}. The dotted line on this plot corresponds to 2\% of the on-disk intensity. The dashed line indicates the cutoff used in the analysis where 45\% of the total intensity is due to stray light, for an assumed stray light intensity equal to 2\% of the on-disk intensity. 
}
\end{figure}

	The intensities for the lines used in the $v_{\mathrm{nt}}$ analysis also show that the scattered light must be $\lesssim 2\%$ of the disk intensity. Figure~\ref{fig:ivsr} shows the intensity versus height of the lines used in our $v_{\mathrm{nt}}$ analysis. In addition, we include line intensity profiles from Fe~\textsc{viii} and Si~\textsc{x}. For each line, the plotted intensity is  that before subtracting off any scattered light contribution. The intensities are normalized to the on-disk intensity $I_{\mathrm{Disk}}$ at about 0.95~$R_{\sun}$. The dotted line in the figure shows the level where $I/I_{\mathrm{Disk}} = 2\%$. The Fe~\textsc{viii} and Si~\textsc{vii} lines fall below this level and the Fe~\textsc{ix} line intensity approaches it at larger heights. This implies that the scattered light level should be about 2\% of the disk intensity or less. The dashed line in the figure indicates the level where scattered light makes up 45\% of the total intensity, assuming the stray light intensity is 2\% of the disk intensity. We consider this level a cutoff in the analysis and do not analyze data where the stray light contamination is larger, because for larger percentages the line width results are sensitive to the scattered light, as is discussed in more detail below. The reason for the different rates of falloff for the various lines is that the plasma is somewhat multithermal. The lines from higher charge states are formed in hotter plasma that has a larger scale height, and therefore the intensity decreases less rapidly than lines formed at lower temperatures \citep{Doschek:ApJ:2001,Hahn:ApJ:2012}. 
 	
\begin{figure}
\centering \includegraphics[width=0.9\textwidth]{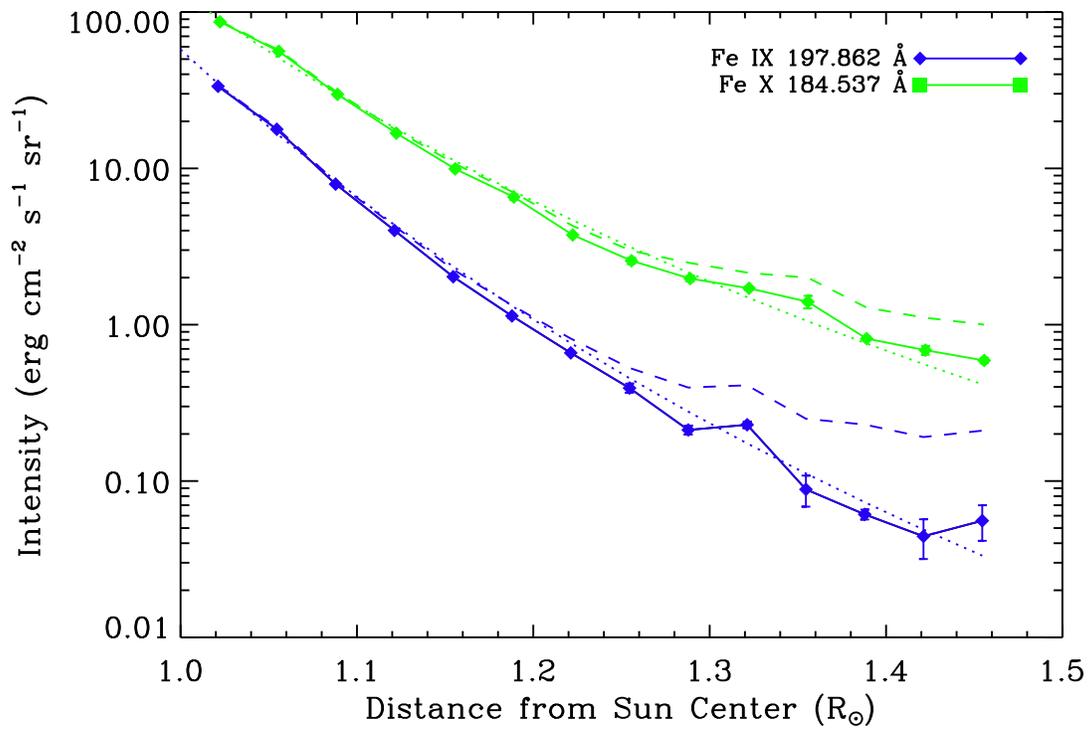}
\caption{\label{fig:subtract} Intensity before stray light subtraction (dashed lines) and after (solid lines). The dotted curve shows a scale height fit to data at low heights where the stray light is less than 20\% of the total intensity. The stray light subtraction brings the intensity profile at large heights into reasonable agreement with the expected scale height falloff. 
}
\end{figure}

 	To demonstrate the effect of stray light subtraction on the intensity, we show in Figure~\ref{fig:subtract} the intensity versus height for lines from Fe~\textsc{ix} and Fe~\textsc{x}. In this figure the solid curve shows the stray-light subtracted intensity, which is used in the analysis, while the dashed line shows the intensity before subtracting stray light. The dotted curve indicates a fit of the data to 
\begin{equation}
I(R)=I(R_0)\exp\left[\frac{-(R-R_0)}{H_{\mathrm{I}}R_{0}R}\right],
\label{eq:hydrointensity}
\end{equation} 	
which describes a scale height falloff with $H_{\mathrm{I}}$ the intensity scale height. This fit was performed for heights where scattered light is less than 20\% of the total intensity, corresponding to $R < 1.19$~$R_{\sun}$ for Fe~\textsc{ix} and $R < 1.29$~$R_{\sun}$ for Fe~\textsc{x}. The figure shows that after we perform the fit with the stray light subtraction, the intensity profile agrees fairly well with the expected scale-height falloff. This further demonstrates the accuracy of the stray light subtraction. 
 	
	Based on the above arguments, we have taken the upper limit for the scattered light level to be 2\% of the on-disk intensity throughout our analysis. This level is also supported by the measurements of \citet{Ugarte:EIS:2010}, who measured stray light during an eclipse where the moon blocked a portion of the solar disk. However, our stray light level is not directly comparable to that of \citet{Ugarte:EIS:2010}. This is because the portion of the disk we observe is near the solar limb and so the on-disk intensity in our data is somewhat increased by limb brightening compared to the \citeauthor{Ugarte:EIS:2010} measurements which were made closer to disk center \citep{Mariska:SolPhys:1975}. We should also note that it is an approximation to use a fixed scattered light value, since the stray light probably decreases with distance above the limb. However, stray light only significantly affects the data at large heights, where the magnitude of the stray light is more important than the variation in it.
	
	Limb brightening introduces some ambiguity about where to measure the stray light intensity. There are several reasons to measure it relative to the lowest on-disk point. First, instrument scattered light can be described as a convolution of emission sources with the point-spread function of the instrument \citep[e.g.,][]{DeForest:ApJ:2009}. The solar disk emission contributes more to the convolution integral since the disk area is larger than the area of the narrow annulus near the peak of the limb-brightening. Another reason for using the on-disk point, is that we have quantified the stray light relative to this position. For example, in Figure~\ref{fig:oxygen} we infer the stray light from the oxygen lines by normalizing to the intensity at 0.95~$R_{\sun}$. It is then consistent to use the same position to estimate the stray light for the line width analysis.
	
\begin{figure}
\centering \includegraphics[width=0.9\textwidth]{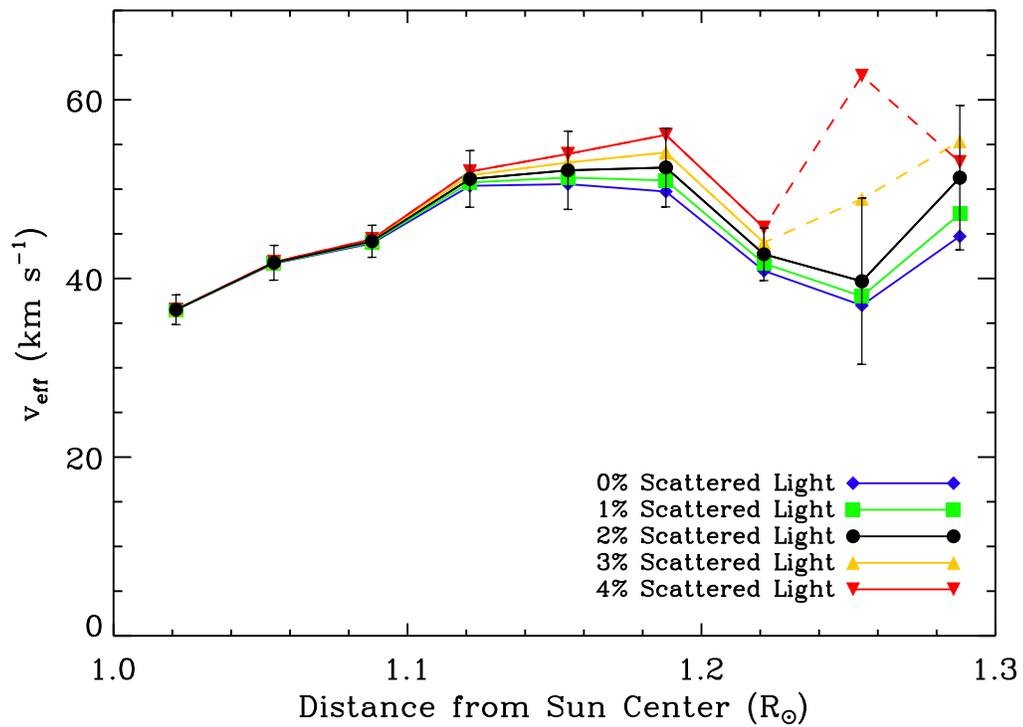}
\caption{\label{fig:fe9stray} Line width $v_{\mathrm{eff}}$ for Fe~\textsc{ix} determined when subtracting different levels of stray light from 0\% to 4\% of the disk intensity. The solid lines connect points where the assumed scattered light is below 45\% of the total intensity and the dashed lines connect points where the stray light contamination is larger. The error bars on the filled circles correspond to the 2\% stray light level used in the analysis.
}
\end{figure}

\begin{figure}
\centering \includegraphics[width=0.9\textwidth]{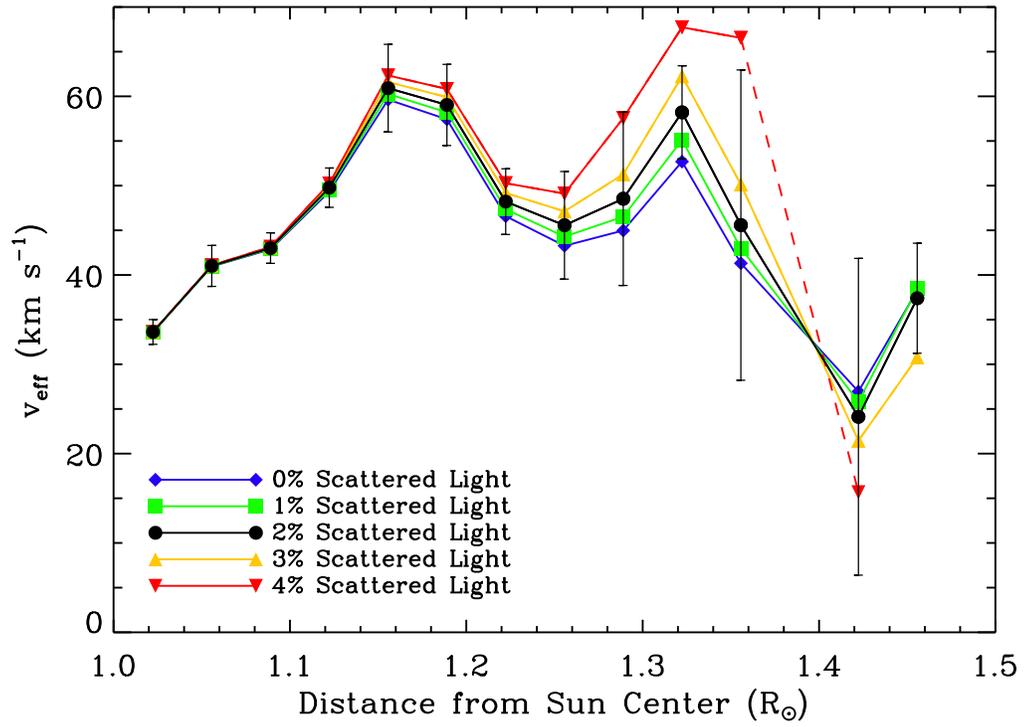}
\caption{\label{fig:fe10stray} Same as Figure~\ref{fig:fe9stray} but for Fe~\textsc{x}
}
\end{figure}

	Furthermore, even if the stray light level were not exactly 2\% of the disk intensity, this would not have a significant effect on our results for the line width. To see the possible effect of stray light on our analysis, we have derived line widths for different levels of stray light. Figures~\ref{fig:fe9stray} and \ref{fig:fe10stray} show the line width $v_{\mathrm{eff}}$ for Fe~\textsc{ix} and Fe~\textsc{x}, respectively. In each case, $v_{\mathrm{eff}}$ is determined after subtracting scattered light having 0, 1, 2, 3, or 4\% of the disk intensity. In these figures the solid lines connect points where stray light makes up less than 45\% of the total observed intensity and the dashed lines connect points that have more stray light contamination. When the stray light level is below 45\%, the various inferred values of $v_{\mathrm{eff}}$ at a given height all lie within the uncertainties, for any stray light intensity from 0\% to 4\% of the disk intensity. This is the reason for applying the 45\% cutoff in the analysis. For very low heights, $R < 1.12$~$R_{\sun}$, the effect of these different stray light levels on $v_{\mathrm{eff}}$ is negligible. This shows that even if the stray light level is different from the 2\% level used in the analysis, the effect on our results is small. 
	
\begin{figure}
\centering \includegraphics[width=0.9\textwidth]{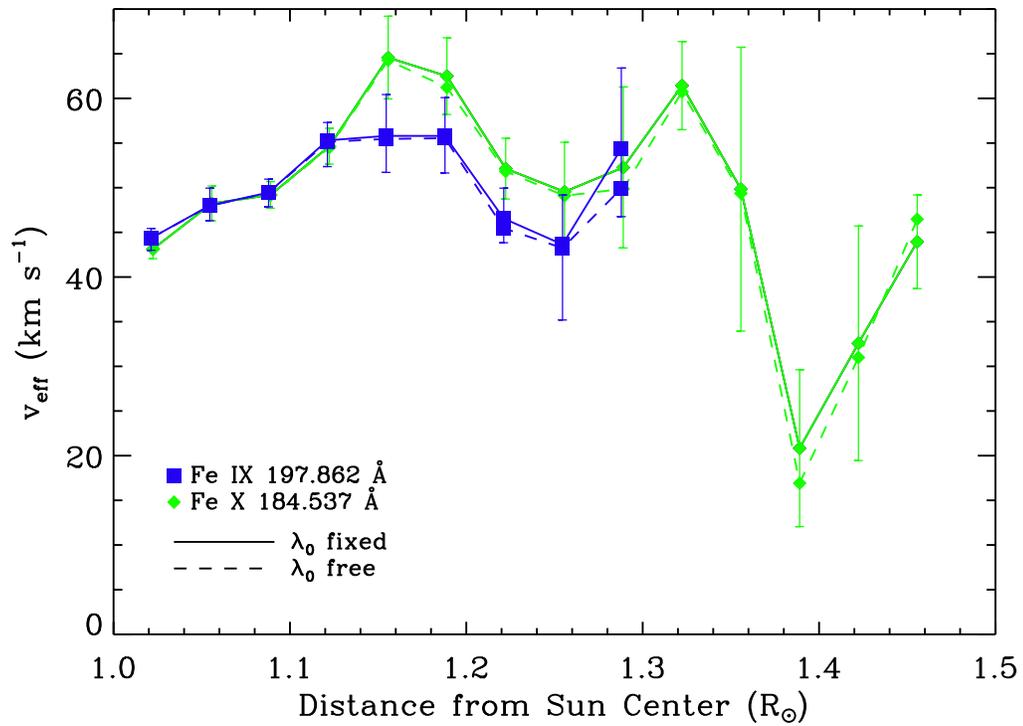}
\caption{\label{fig:centroideffect} Line width for Fe~\textsc{ix} and Fe~\textsc{x} lines using a fixed stray light centroid $\lambda_{0}$ position (solid line) or allowing it to vary as parameter in the fit (dashed line). The error bars represent the uncertainties for the fixed $\lambda_{0}$ case. Allowing the stray light $\lambda_{0}$ to vary has no significant effect on the line width.
}
\end{figure}

	One other systematic effect of the stray light subtraction that we can readily check is the influence of the stray light centroid position. For the above analysis we fixed the stray light centroid $\lambda_{0}$ to the value measured on the disk. This seems the most reasonable, since it is expected that stray light comes from the bright solar disk. However, we can also allow the centroid to vary freely to see if this would have an effect on the inferred $v_{\mathrm{eff}}$. Figure~\ref{fig:centroideffect} shows $v_{\mathrm{eff}}$ for the Fe~\textsc{ix} and Fe~\textsc{x} lines for using either fixed $\lambda_{0}$ or allowing $\lambda_{0}$ to vary as a parameter of the fit. The difference between $v_{\mathrm{eff}}$ for the two cases is well within the fitting uncertainties. 	
 	
 	The above analysis characterizes the scattered light in our EIS observations. First, we find that at large heights above the disk the scattered light intensity is very low, and is below 2\% of the intensity at the lowest on-disk point in our data. This implies that the stray light contribution must be even smaller. Second, we have found that as long as scattered light makes up less than about 45\% of the total intensity, the inferred line widths do not change significantly for a substantial range of different stray light intensities. Finally, we have shown that if the centroid position of the scattered light line profiles are allowed to vary, the inferred line widths remain the same. Thus, with a few constraints, our results are insensitive to scattered light. 
		
\section{Summary} \label{sec:sum}
		
		We have found that Alfv\'en waves in a polar coronal hole possess sufficient energy to heat the coronal hole and that this energy is actually dissipated from the waves at sufficiently low heights to heat the corona. To show this we determined separately the thermal and non-thermal components of spectral line broadening in a coronal hole. Our method relies on the observation that waves are undamped at very low heights and on the assumption that the temperature of each ion does not change with height at low heights. From the derived $v_{\mathrm{nt}}$ we show that the energy carried by the waves is $6.7 \pm 0.7 \times10^{5}$~$\mathrm{erg\,cm^{-2}\,s^{-1}}$, which is sufficient to heat the coronal hole and accelerate the fast solar wind. About 85\% of this initial energy is damped by 1.44~$R_{\sun}$. The length scale for the damping is about $0.18\pm0.04$~$R_{\sun}$, with a corresponding timescale of about $68\pm15$~s. Although our measurements are limited to $R < 1.5$~$R_{\sun}$, they suggest that enough energy remains in the waves to provide the extended heating of the solar wind above the sonic point that models show is required to accelerate the fast solar wind to the speeds observed far from the Sun. Additionally, we measured $T_{\mathrm{i}}$ for each ion to be in the range of about 1 - 2 MK. We found a weak trend where low $q/M < 0.2$ ions have the highest temperature, $q/M \approx 0.2$ - 0.25 are lower with $T_{\mathrm{i}}\approx T_{\mathrm{e}}$, and $q/M > 0.25$ have a slightly increasing temperature. Our uncertainties are too large to distinguish between the predictions of two ion heating models. Those models are also not realistic for these heights since they neglect Coulomb collisions. Our results, though, do demonstrate that such a comparison is possible in principle, needing only additions to the model and higher quality observational data.
	
\acknowledgements
We thank Leon Ofman for helpful discussions. This work was supported in part by the NASA Solar Heliospheric Physics 
program grant NNX09AB25G and the NSF Division of Atmospheric and Geospace Sciences SHINE program
grant AGS-1060194.

 	

\begin{deluxetable}{llccccc}
\tabletypesize{\footnotesize}
\tablecaption{Line List. \label{table:linelist}}
\tablewidth{0pt}
\tablecolumns{7}
\tablehead{
	\colhead{~} & 
	\colhead{Ion} & 
	\colhead{~} & 
	\colhead{$\lambda$ (\AA)\tablenotemark{1}} & 
	\multicolumn{3}{c}{Used for:} \\
	
	\colhead{} & \colhead{} & \colhead{} & \colhead{} & 
	\colhead{Eq.~(\ref{eq:ls2}) Fit} & 
	\colhead{$v_{\mathrm{nt}}(R)$} & 
	\colhead{$T_{\mathrm{i}}(1.05 R_{\sun})$}
	}
\startdata
&O \textsc{vi} & &183.937 	& & & $\ast$ \\
&O \textsc{vi} & &184.118 	& & & $\ast$ \\
&Mg \textsc{vii} & &276.154 & & & $\ast$ \\
&Si \textsc{vii} & &272.648 & $\ast$ & & $\ast$ \\
&Si \textsc{vii} & &275.361 & $\ast$ & $\ast$ & $\ast$ \\
&Si \textsc{vii} & &275.676 & $\ast$ & & $\ast$ \\
&Si \textsc{x} & &258.374 	&	& & $\ast$ \\
&Si \textsc{x} & &261.057 	& &	& $\ast$ \\
&Si \textsc{x} & &271.992 	& & & $\ast$ \\
&Si \textsc{x} & &277.264 	& & & $\ast$ \\
&S  \textsc{x} & &264.231 	& & & $\ast$ \\
&Fe \textsc{viii} & &185.213 & $\ast$ & & $\ast$ \\
&Fe \textsc{viii} & & 186.599 & $\ast$ & & $\ast$ \\
&Fe \textsc{viii} & & 194.661 & $\ast$ & & $\ast$ \\
&Fe \textsc{ix} & &188.497 & $\ast$ & & $\ast$ \\
&Fe \textsc{ix} & &189.941 & $\ast$ & & $\ast$ \\
&Fe \textsc{ix} & &197.862 & $\ast$ & $\ast$ & $\ast$ \\
&Fe \textsc{x} & &184.537 & $\ast$ & $\ast$ & $\ast$ \\
&Fe \textsc{x} & &190.037 & & & $\ast$ \\
&Fe \textsc{x} & &193.715 & & & $\ast$ \\
						&	&	& 257.259 & & & $\ast$ \\*
&\raisebox{2.5ex}[0pt]{Fe \textsc{x}}&\raisebox{2.5ex}[0pt]{$\Big\{$} & 257.263 & & & $\ast$ \\
&Fe \textsc{xi} & &180.401 & & & $\ast$ \\ 
&Fe \textsc{xi} & &188.217 & $\ast$ & $\ast$ & $\ast$ \\
&Fe \textsc{xi} & &188.299 & & & \nodata \tablenotemark{2} \\
\enddata
\tablenotetext{1}{Wavelengths from CHIANTI \citep{Dere:AA:1997, Landi:ApJ:2012}.}
\tablenotetext{2}{$\Delta \lambda$ was constrained to be identical for Fe~\textsc{xi} 188.217~\AA\ and 188.299~\AA. }
\tablecomments{Brackets indicate blends from the same ion.}
\end{deluxetable}

\begin{deluxetable}{lll}
\tablecolumns{3}
\tablewidth{0pc}
\tablecaption{Values for $v_{\mathrm{th}}$ and $v_{\mathrm{nt}}$ at 1.05~$R_{\sun}$ from fitting Equation~(\ref{eq:ls2}) over 1.02 - 1.12~$R_{\sun}$.
\label{table:fitresults}}
\tablehead{
	\colhead{Ion} & 
	\colhead{$v_{\mathrm{th}}$~$(\mathrm{km\,s^{-1}})$ } & 
	\colhead{$v_{\mathrm{nt}}$~$(\mathrm{km\,s^{-1}})$ } 
}
\startdata
Si~\textsc{vii} & $23.5\pm1.5$ & $33.6 \pm 1.2$ \\
Fe~\textsc{viii} & $19.9\pm2.7$ & $29.8 \pm 1.8$ \\
Fe~\textsc{ix} & $20.4 \pm 3.0$ & $34.9 \pm 1.8$ \\
Fe~\textsc{x} & $18.7 \pm 2.6$ & $34.5 \pm 1.5$ \\
Fe~\textsc{xi} & $25.8 \pm 5.4$ & $32.2 \pm 4.2$
\enddata
\end{deluxetable}

\begin{deluxetable}{ccc}
\tablecolumns{3}
\tablewidth{0pc}
\tablecaption{Non-thermal Velocity and Energy Flux Density. 
\label{table:ergresults}}
\tablehead{
	\colhead{R $(R_{\sun})$} & 
	\colhead{$F$~$(10^{5}\,\mathrm{erg\,cm^{-2}\,s^{-1}})$} & 
	\colhead{$F\frac{A(R)}{A(R_{\sun})}$~$(10^{5}\,\mathrm{erg\,cm^{-2}\,s^{-1}})$}
}
\startdata
1.02 & $5.5\phn \pm 1.3$ & $6.1\phn \pm 1.4$ \\
1.05 & $5.6\phn \pm 1.1$ & $7.1\phn \pm 1.4$ \\
1.08 & $4.53 \pm 0.91$ & $6.5\phn \pm 1.3$ \\
1.11 & $4.45 \pm 0.93$ & $7.3\phn \pm 1.5$ \\
1.14 & $3.75 \pm 0.93$ & $6.9\phn \pm 1.7$ \\
1.17 & $2.30 \pm 0.57$ & $4.7\phn \pm 1.2$ \\
1.20 & $2.14 \pm 0.85$ & $ 4.9\phn \pm 1.9$ \\
1.23 & $1.25 \pm 0.29$ & $3.18 \pm 0.73$ \\
1.26 & $0.91 \pm 0.39$ & $2.6\phn \pm 1.1$ \\
1.29 & $0.89 \pm 0.33$ & $2.8\phn \pm 1.0$ \\
1.32 & $0.64 \pm 0.16$ & $2.20 \pm 0.57$ \\
1.35 & $0.65 \pm 0.42$ & $2.5\phn \pm 1.6$ \\
1.38 & $0.48 \pm 0.16$ & $1.98 \pm 0.68$ \\
1.41 & $0.34 \pm 0.33$ & $1.5\phn \pm 1.5$ \\
1.44 & $0.20 \pm 0.10$ & $0.98 \pm 0.50$
\enddata
\end{deluxetable}

\begin{deluxetable}{lccc}
\tablecolumns{3}
\tablewidth{0pc}
\tablecaption{Ion Temperatures.
\label{table:Tiresults}}
\tablehead{
	\colhead{} &
	\colhead{Ion} & 
	\colhead{$q/M$~$(\frac{e}{\mathrm{amu}})$} & 
	\colhead{$T_{\mathrm{i}}$~(MK)}
}
\startdata
&O~\textsc{vi} & 0.31 & $2.9\phn \pm 1.2$ \\
&Mg~\textsc{vii} & 0.25 & $0.91 \pm 0.44$ \\
&Si~\textsc{vii} & 0.21 & $1.04 \pm 0.47$ \\
&Si~\textsc{x} & 0.32 & $1.41 \pm 0.45$ \\
&S~\textsc{x} & 0.28 & $1.44 \pm 0.57$ \\ 
&Fe~\textsc{viii} & 0.13 & $1.91 \pm 0.74$ \\
&Fe~\textsc{ix} & 0.14 & $2.32 \pm 0.86$ \\
&Fe~\textsc{x} & 0.16 & $1.50 \pm 0.96$ \\
&Fe~\textsc{xi} & 0.18 & $2.74 \pm 0.94$
\enddata
\end{deluxetable}

\bibliography{Damping}
\end{document}